\documentclass [12pt,a4paper]{article}

\usepackage{isridef}  % from QFT-PRO.20

\newcommand{\HU}{\text{H}} % Heaviside unit step function (FOR "EJP" papers!)

\def\a{\underline{a}}         % Sequence in a
\def\limea{\lim_{\substack{
                \epsilon \rightarrow 0\\
                       a \rightarrow 0}}}
\def\lime{\lim_{\epsilon \rightarrow 0}}

\numberwithin{equation}{section}

\begin{document}

\title{\bf\vspace{-3.5cm} The classical point-electron in the
                sequence algebra $(\mathcal{C}^\infty)^\mathcal{I}$}
%               ====================================================

\author{
         {\bf Andre Gsponer}\\
         {\it Independent Scientific Research Institute}\\ 
         {\it Oxford, OX4 4YS, England}
       }

\date{ISRI-08-03.10 ~~ \today}

\maketitle

\begin{abstract}

In arXiv:0806.4682 the self-energy and self-angular momentum (i.e., electromagnetic mass and spin) of a classical point-electron were calculated in a Colombeau algebra.  In the present paper these quantities are calculated in the better known framework of `regularized distributions,' i.e., the customary setting used in field-theory to manipulate diverging integrals, distributions, and their products.  The purpose is to compare these two frameworks, and to highlight the reasons why the Colombeau theory of nonlinear generalized functions could be the physically preferred setting for making these calculations.  In particular, it is shown that, in the Colombeau algebra, the point-electron's mass and spin are \emph{exact} integrals of squares of delta-functions, whereas this is only an approximation in the customary framework.

%`regularized distributions,' =  `distribution representatives'

\end{abstract}	

%%%%%%%%%%%%%%%%%%% BEGIN WORK %%%%%%%%%%%%%%%%%%%%%%%%%
%\newpage

%%%%%%%%%%%%%%%%%%% END WORK %%%%%%%%%%%%%%%%%%%%%%%%%%%

\section{Introduction} 
%=====================
\label{int:0}

In Reference \cite{GSPON2008B} we calculated the electromagnetic mass and spin of a classical point-electron defined as a pole-dipole singularity of the Maxwell field.  Since these quantities are nonlinear in the fields, as well as divergent, they were calculated in a Colombeau algebra which provided the framework required to multiply the fields interpreted as distributions in a mathematically meaningful manner.

   The Colombeau algebra is however only one possible associative, commutative differential algebra of generalized functions containing the space $\mathcal{D}'$ of Schwartz distributions as a linear subspace:  The distinctive feature of the Colombeau algebra $\mathcal{G}$ is that the embedding of $\mathcal{D}'$ is such that $\mathcal{C}^\infty$, the algebra of smooth functions, is a faithful differential subalgebra of $\mathcal{G}$, see \cite{COLOM1984-,COLOM1985-}.  This feature has the momentous consequence that \emph{all properties of a physical theory that are valid for smooth functions remain valid when this theory is embedded into a Colombeau algebra}.  This is for instance the case of classical electrodynamics, which is essentially a continuum theory:  Embedding classical electrodynamics into a Colombeau algebra leads to a theory that is mathematically meaningful and consistent for both smooth and distributional potentials, fields, and charge-current distributions, as well as any of their linear or nonlinear algebraic combinations.

   Yet, not much use was made of the powerful properties of the Colombeau algebra in the calculations made in Reference \cite{GSPON2008B}:  Apart from providing the framework allowing distributions to be multiplied, its main application was to define a generalized function $\UPS$ which enabled to greatly simplify the computation of nonlinear quantities such as the self-energy and the self-angular-momentum.  In fact, the $\mathcal{G}$-setting would have been truly essential if these nonlinear quantities were used in subsequent calculations in which they would have been further multiplied by other nonlinear generalized functions.  Instead, they were simply integrated to yield scalars quantities, i.e., mass and spin.

   In this paper it will be shown that the electromagnetic mass and spin, as well as the total self-force and the total self-momentum, can actually be calculated in the more elementary framework of `regularized distributions,' that is in the `sequence algebra' which is possibly more familiar to physicists than the Colombeau algebra.\footnote{For an elementary introduction to Colombeau algebras and a selection of references see \cite{GSPON2006B} or \cite{GSPON2008A}.  For a more comprehensive introduction see \cite{COLOM1992-}.}

  This will enable to compare the mathematically rigorous calculations made in Reference \cite{GSPON2008B} to the analogous calculations made in this paper using the empirically justified methods well known to physicists, and thus to highlight the differences between these two methodologies in order to clarify the conceptual advantages of the Colombeau method.  This is the main objective of this paper.

\section{The sequence algebra $(\mathcal{C}^\infty)^\mathcal{I}$} 
%================================================================
\label{seq:0}

Let us consider the set $(\mathcal{C}^\infty)^\mathbb{N}$ of all the usual sequences of smooth functions on $\mathbb{R}^n$. As is well known, when considered with the usual termwise operations on sequences of functions, $(\mathcal{C}^\infty)^\mathbb{N}$ is a commutative differential algebra.  Namely, its elements can be added and multiplied arbitrarily, and they can be differentiated any number of times.  Now, to be formally as close as possible to the definition of the Colombeau algebra given in Ref.~\cite{GSPON2008B}, we will use the term `sequence' for expressions such as $\lim_{\epsilon \rightarrow 0} F_\epsilon$ where $\epsilon \in \mathcal{I}$, with $\mathcal{I} \DEF ]0,1[$, even though mathematicians reserve this term to mappings $n \mapsto F_n$ where $n \in \mathbb{N}$, so that $\epsilon \rightarrow 0$ corresponds to $1/n \rightarrow 0$ as $n \rightarrow \infty$.  Then, instead of $(\mathcal{C}^\infty)^\mathbb{N}$, we designate the `sequence algebra' by $(\mathcal{C}^\infty)^\mathcal{I}$, and we define:

\begin{definition}[Sequence algebra] 
%..................................
\label{seq:defi:1}
Let $\Omega$ be an open set in $\mathbb{R}^n$, let $\mathcal{I} = ]0,1[ \subset \mathbb{R}$, and let $\epsilon \in \mathcal{I}$ be a parameter.  The `sequence algebra' is the differential algebra 
\begin{align}
\nonumber
    (\mathcal{C}^\infty)^\mathcal{I}(\Omega) \DEF  \Bigl\{
     f_{\epsilon} : ~~
     \mathcal{I} \times \Omega &\rightarrow \mathbb{C},\\
\label{seq:1}
                             (\epsilon, \vec{x}\,) &\mapsto 
                                        f_{\epsilon}(\vec{x}\,) \Bigr\},
\end{align}
where the sequencies $f_{\epsilon}$ are $\mathcal{C}^\infty$ functions in the variable $\vec{x} \in \Omega$.  The compactly supported distributions are embedded in $(\mathcal{C}^\infty)^\mathcal{I}$ by convolution with the scaled regularizer $\rho_\epsilon$, i.e.,\footnote{This definition due to Colombeau differs by a sign from the usual definition of regularization.}
\begin{align}
\label{seq:2}
  f_\epsilon(\vec{x}\,) \DEF  \int_\Omega \frac{dy^n}{\epsilon^n}
            ~\rho\Bigl(\frac{\vec{y}-\vec{x}}{\epsilon}\Bigr) ~f(\vec{y}\,)
         = \int_\Omega dz^n~\rho(\vec{z}\,) ~f(\vec{x} + \epsilon \vec{z}\,),
\end{align}
where the regularizing function is written $\rho$ to make clear that it is not a Colombeau mollifier $\eta  \in \mathcal{A}_\infty$.  That is, the only constraint on the regularizer $\rho \in \mathcal{S}$ is to be normalized in such a way that
\begin{equation} \label{seq:3}
    \int_\Omega dz^n~\rho(\vec{z}\,) = 1.
\end{equation}
\end{definition}
For example, the regularizations of the Dirac and Heaviside functions in $\Omega = \mathbb{R}$, that is their embeddings in $(\mathcal{C}^\infty)^\mathcal{I}(\mathbb{R})$, are
\begin{equation} \label{seq:4}
     \delta_\epsilon(x) 
      = \frac{1}{\epsilon} \rho\Bigl(-\frac{x}{\epsilon}\Bigr),
    \qquad \text{and} \qquad
 \HU_\epsilon(x) = \int_{-x/\epsilon}^{\infty} dz~\rho(z)
                 = \int_{-\infty}^{x/\epsilon} dz~\rho(-z).
\end{equation}

   Conversely it is possible to recover any distribution $f$ from its regularizations by means of its definition as a functional, i.e., as the equivalence class
\begin{equation}
\label{seq:5}
        f(T) \DEF \lim_{\epsilon \rightarrow 0} 
                 \int_{-\infty}^{+\infty} dx~f_\epsilon(x) ~T(x),
                \qquad \forall T(x) \in \mathcal{D},
\end{equation}
where $T$ is any test-function, and  $f_\epsilon$ any representative sequence of $f$.  But, of course, not all sequences are distributions.  For example, the square of Dirac's $\delta$-function defined as the square of its embedding \eqref{seq:4}, i.e.,
\begin{equation} \label{seq:6}
     (\delta^2)_\epsilon(x) \DEF (\delta_\epsilon)^2(x)  
      = \frac{1}{\epsilon^2} \rho^2\Bigl(-\frac{x}{\epsilon}\Bigr),
\end{equation}
is not a distribution because, making the change of variables $x=-\epsilon y$,
\begin{equation} \label{seq:7}
    \lim_{\epsilon \rightarrow 0} \frac{1}{\epsilon}
                 \int_{-\infty}^{+\infty} \frac{dx}{\epsilon}
            \rho^2\Bigl(\frac{-x}{\epsilon}\Bigr) T(x)
    = T(0) \lim_{\epsilon \rightarrow 0} \frac{M[^2_0]}{\epsilon}
    = \infty,
\end{equation}
where the constant $M[^2_0]$ is the moment
\begin{equation} \label{seq:8}
            M[^2_n] \DEF \int_{-\infty}^{+\infty} dy~ y^n \, \rho^2(y).
\end{equation}

  The sequence algebra $(\mathcal{C}^\infty)^\mathcal{I}$ contains the distributions embedded according to \eqref{seq:2} as a subspace.  These embedded distributions can therefore be `multiplied' in the sense that their product is associative and commutative in $(\mathcal{C}^\infty)^\mathcal{I}$.  But apart from being the product of two regularized objects, this product has \emph{no} particular meaning:  In general, the product of two embedded distributions or continuous functions will not be a distribution or continuous function, and even the product of two embedded smooth functions will \emph{not} be equal to the embedding of the product of the two original smooth functions.

    On the other hand, and especially in the perspective of its physical applications, the remarkable property of the Colombeau algebra $\mathcal{G}$ is that the algebra $\mathcal{C}^\infty$ of the smooth functions is a faithful differential subalgebra of $\mathcal{G}$, thanks to the Colombeau regularization which insures the compatibility between the products of these functions in $\mathcal{C}^\infty$ and the products of their embeddings in $\mathcal{G}$.  It then follows that the properties which in a physical theory are true for $\mathcal{C}^\infty$ objects are automatically extended to their  $\mathcal{G}$-embeddings when that theory is considered in $\mathcal{G}$ rather then in $\mathcal{C}^\infty$.

   However, working in $(\mathcal{C}^\infty)^\mathcal{I}$ can nevertheless lead to physically meaningful results, and it is of interest in view of getting a better understanding of the physical significance of the Colombeau algebra to compare calculations made in $(\mathcal{C}^\infty)^\mathcal{I}$ to those made in $\mathcal{G}$, as will be done in this paper.

\section{Point singularities in $\mathcal{D}'$ and in $(\mathcal{C}^\infty)^\mathcal{I}$}
%===================================================================
\label{poi:0}

   While classical electrodynamics is fundamentally a continuum theory, it is possible to consistently introduce point charges through distribution theory. The basic idea is to replace the classical Coulomb potential $e/r$ of a point charge by the weak limit of the sequence of distributions \cite[p.\,144]{TEMPL1953-}, \cite[p.\,51]{SCHUC1991-},
\begin{equation} \label{poi:1}
  \phi(r) \DEF \frac{e}{r} \HU_{\a}(r),
\qquad \text{where} \qquad
             \HU_{\a}(r) \DEF \lim_{a \rightarrow 0} \HU(r-a),
\end{equation}
where $e$ is the electric charge of an electron at rest at the origin of a polar coordinate system, and $r = |\vec{r}\,|$ the modulus of the radius vector.\footnote{We write $\a$ rather than $a$ in subscript to emphasize that $a$ is not an index but the parameter of a limiting sequence.  This parameter $a$ is also not a regularizing parameter such as $\epsilon$ in \eqref{seq:2}.}  Consistent with Schwartz's local structure theorem, $\phi(r)$ is the derivative of $e\lim_{a \rightarrow 0} \log(r/a) \HU(r-a)$, a $\mathcal{C}^0$ function $\forall r \geq 0$.  The infinitesimal cut-off $a > 0$ insures that $\phi(r)$ is a well defined piecewise continuous function for all $r \geq 0$, whereas the classical Coulomb potential $e/r$ is defined only for $r > 0$.  It is then readily verified, using \eqref{seq:5}, that \eqref{poi:1} is a distribution.  Indeed,
\begin{equation} \label{poi:2}
   \forall T \in \mathcal{D}, \qquad 
   \iiint_{\mathbb{R}^3} d^3r~ \phi(r) T(r) =
  4\pi e \int_a^\infty dr~ r T(r) \quad \in \mathbb{R},
\end{equation}
because $T \in \mathbb{D}$ has compact support so that the integral is bounded.  

   Differentiating \eqref{poi:2} in the sense of distributions, as is done for example in Ref.\,\cite[p.\,144]{TEMPL1953-}, one can calculate the distributional field $\vec{E} = -\vec{\nabla} \phi$ and charge density $\varrho = \vec{\nabla} \cdot \vec{E}$.

  But in order to show how this is done in $(\mathcal{C}^\infty)^\mathcal{I}$, and how the resulting $(\mathcal{C}^\infty)^\mathcal{I}$-functions relate to the corresponding distributions, we begin by regularizing $\phi$ according to \eqref{seq:2}, so that the regularized Coulomb potential is 
\begin{equation} \label{poi:3}
  \phi_\epsilon(r) = \bigl(\frac{e}{r} \HU_{\a}\bigr)_\epsilon(r)
             = e \lim_{a \rightarrow 0}
                 \int_{\frac{a-r}{\epsilon}}^\infty dy~
                 \frac{\rho(y)}{r+\epsilon y},
                 \qquad \forall r \geq 0.
\end{equation}
One then easily verifies that the distributional Coulomb potential \eqref{poi:1} is recovered when $\epsilon \rightarrow 0$.  Indeed, in that limit, \eqref{poi:3} tends towards $0$ for $r < a$, and towards $1$ for $r > a$.  Thus
\begin{equation} \label{poi:4}
  \phi_\epsilon(r) \ASS \frac{e}{r} \HU_{\a}(r) = \phi(r),
\end{equation}
which reverts to the classical Coulomb potential $e/r$ as $a \rightarrow 0$.

  Calculating the regularized Coulomb field is now straightforward because the regularized potential $(e\HU_{\a}/r)_\epsilon$ is $\mathcal{C}^\infty$ in the variable $r$.  It comes
\begin{equation} \label{poi:5}
  \vec{E}_\epsilon(\vec{r}\,) = -\vec{\nabla} \phi_\epsilon(r)
              = e \lim_{a \rightarrow 0} \Bigl(
            \int_{\frac{a-r}{\epsilon}}^\infty dy~
                            \frac{\rho(y)}{(r+\epsilon y)^2}
          - \frac{1}{\epsilon a}\rho\bigl(\frac{a -r}{\epsilon}\bigr)
                        \Bigr) \vec{u},
\end{equation}
where $\vec{u} = \vec{\nabla} r$ is the unit vector in the direction of $\vec{r}$.  Introducing the notation
\begin{equation} \label{poi:6}
   \delta_{\a}(r) \DEF \lim_{a \rightarrow 0} \delta(r-a),
   \qquad \text{so that} \qquad
   \lim_{a \rightarrow 0}
   \frac{1}{\epsilon a}\rho\bigl(\frac{a -r}{\epsilon}\bigr)
 = \bigl(\frac{1}{a} \DUP_{\a}\bigr)_\epsilon(r),
\end{equation}
this electric field can be written in the more convenient form  \cite{GSPON2006B}
\begin{equation} \label{poi:7}
  \vec{E}_\epsilon(\vec{r}\,) = e \Bigl(
            \bigl(\frac{1}{r^2} \HU_{\a}\bigr)_\epsilon(r)
          - \bigl(\frac{1}{a} \DUP_{\a}\bigr)_\epsilon(r)
                        \Bigr) \vec{u}.
\end{equation}
By an appeal to test functions we easily verify that the field $\vec{E}_\epsilon$ is a distribution, and that the $\DUP$-function in \eqref{poi:7} gives a nul contribution when evaluated on a test function.  Thus
\begin{equation} \label{poi:8}
  \vec{E}_\epsilon(\vec{r}\,) \ASS \frac{e}{r^2} \HU_{\a}(r) \vec{u}
                          \DEF \vec{E}(\vec{r}\,),
\end{equation}
where $\vec{E}(\vec{r}\,)$ is the distributional Coulomb field which in the limit $a \rightarrow 0$ yields the classical Coulomb field $e\vec{r}/r^3$.  Therefore, the distribution $\vec{E}(\vec{r}\,)$ associated to the regularized field $\vec{E}_\epsilon$ does not contain the $\DUP$-function contribution on the right of \eqref{poi:7}.

  To get the Coulomb charge density we have to calculate the divergence of \eqref{poi:5}.  In standard distribution theory one would then ignore the term on the right because it corresponds to a $\DUP$-function which, as we have just seen, gives no contribution when evaluated on a test function.  However, in $(\mathcal{C}^\infty)^\mathcal{I}$, this term cannot be ignored if we subsequently calculate quantities in which $\vec{E}_\epsilon$, or any of its derivatives, is a factor in a product.  Calculating $\varrho_\epsilon$ is therefore somewhat laborious, but still elementary. It yields, using $\vec{\nabla} \cdot \vec{u} = 2/r$,
\begin{align}
\nonumber
   4\pi \varrho_\epsilon(r) &= \vec{\nabla} \cdot \vec{E}_\epsilon(\vec{r}\,)
              = e \lim_{a \rightarrow 0} \Bigl( ~
        \frac{2}{r} \int_{\frac{a-r}{\epsilon}}^\infty dy~
                            \frac{\rho(y)}{(r+\epsilon y)^2}
        - \frac{2}{\epsilon a r}\rho\bigl(\frac{a -r}{\epsilon}\bigr)\\
 \label{poi:9}
        &+ \frac{1}{\epsilon a^2}\rho\bigl(\frac{a -r}{\epsilon}\bigr)
        -2 \int_{\frac{a-r}{\epsilon}}^\infty dy~
                            \frac{\rho(y)}{(r+\epsilon y)^3}
        + \frac{1}{\epsilon^2 a}\rho'\bigl(\frac{a -r}{\epsilon}\bigr)
                       ~ \Bigr),
\end{align}
which can be rewritten in the less cumbersome form
\begin{align}
\nonumber
   4\pi \varrho_\epsilon(r) &= e \lim_{a \rightarrow 0} \Bigl( ~
        \frac{2}{r} \bigl(\frac{1}{r^2} \HU_{\a}\bigr)_\epsilon(r)
                 -2 \bigl(\frac{1}{r^3} \HU_{\a}\bigr)_\epsilon(r)\\
 \label{poi:10}
        &+ \bigl(\frac{1}{a^2} \DUP_{\a}\bigr)_\epsilon(r)
         - \frac{2}{r}\bigl(\frac{1}{a} \DUP_{\a}\bigr)_\epsilon(r)
         - \bigl(\frac{1}{a}  \DUP'_{\a}\bigr)_\epsilon(r)
                       ~ \Bigr).
\end{align}
The distribution associated to this expression is obtained by evaluating it on a test function \cite[Sec.\,8]{GSPON2006B}, i.e., 
\begin{equation} \label{poi:11}
  \varrho_\epsilon(r) \ASS \frac{e}{4\pi r^2} \DUP_{\a}(r) \DEF \varrho(r),
\end{equation}
which yields the classical point-charge density $ e\delta^3(r) = e \delta(r)/4\pi r^2$ as $a \rightarrow 0$.

  Finally, generalizing definitions \eqref{poi:1} and \eqref{poi:3}, we note that the cut-off $a$ implies the identities, 
\begin{equation} \label{poi:12}
   \frac{1}{r^n} \HU_{\a}(0) = 0,
   \qquad \text{and} \qquad
   \lim_{\epsilon \rightarrow 0} 
   \bigl(\frac{1}{r^n} \HU_{\a}\bigr)_\epsilon(0) = 0,
\end{equation}
which are valid $\forall n \in \mathbb{Z}$, and which will be useful in what follows.

\section{Electric monopole: Self-energy}
%=======================================
\label{sen:0}

In the classical theory, where $\vec{E}(\vec{r}\,)$ is just the Coulomb field $e\vec{r}/r^3$, the self-energy of a point charge is the integral
\begin{equation} \label{sen:1}
 U_{\text{ele}} \DEF \frac{1}{8\pi} \iiint_{\mathbb{R}^3} d^3r~ \vec{E}^2 
                    = \frac{1}{2} \int_0^\infty dr~ r^2\frac{e^2}{r^4}
                    = \frac{e^2}{2}\lim_{r \rightarrow 0} \frac{1}{r} = \infty.
\end{equation}

    In distribution theory the Coulomb field is the distribution $\vec{E}(\vec{r}\,) = e\HU_{\a}\vec{r}/r^3$ defined by \eqref{poi:8}.  Then, apart from expressing the self-energy $U_{\text{ele}}(T)$ as a function of the parameter $a$, we still have the same divergent result
\begin{equation} \label{sen:2}
 U_{\text{ele}}(1) = \frac{1}{8\pi}  \BRA \vec{E}^2 | 1 \KET
                    = \frac{e^2}{2}\lim_{a \rightarrow 0} \frac{1}{a} = \infty,
\end{equation}
even if $\vec{E}^2$ is evaluated on a test-function $T\neq 1$.

    We now calculate the self-energy in $(\mathcal{C}^\infty)^\mathcal{I}$.  With the Coulomb field expressed as \eqref{poi:7} we have therefore to integrate\footnote{To further simplify the notation we leave implicit the $r$-dependence of $\HU_{\a}$ and $\DUP_{\a}$.}
\begin{equation} \label{sen:3}
 U_{\text{ele}}   = \lim_{\epsilon \rightarrow 0}
        \frac{e^2}{2} \int_0^\infty dr~ r^2 \Bigl(
        \bigl(\frac{1}{r^2} \HU_{\a}\bigr)_\epsilon^2
      -2\bigl(\frac{1}{r^2} \HU_{\a}\bigr)_\epsilon
        \bigl(\frac{1}{a} \DUP_{\a}\bigr)_\epsilon
      + \bigl(\frac{1}{a} \DUP_{\a}\bigr)_\epsilon^2
          \Bigr),
\end{equation}
where the explicit forms of the representatives are given by \eqref{poi:5}, and where the limit $\epsilon \rightarrow 0$ has to be taken before the limit $a \rightarrow 0$, which is implicit in the symbols $\HU_{\a}$ and $\DUP_{\a}$.  However, as a quite similar expression will have to be integrated when calculating the self-force, we consider the more general integral
\begin{equation} \label{sen:4}
     \mathcal{M}_n(\a,\epsilon)   =  \int_0^\infty dr~ r^n \Bigl(
        \bigl(\frac{1}{r^2} \HU_{\a}\bigr)_\epsilon^2
      -2\bigl(\frac{1}{r^2} \HU_{\a}\bigr)_\epsilon
        \bigl(\frac{1}{a} \DUP_{\a}\bigr)_\epsilon
      + \bigl(\frac{1}{a} \DUP_{\a}\bigr)_\epsilon^2
          \Bigr),
\end{equation}
which multiplied by $\lim_{\epsilon \rightarrow 0} e^2/2$ is the $n$-th moment of the energy-density $\vec{E}^2/4\pi$.

\noindent (i) The integral of the $\HU_{\a}^2$ term in \eqref{sen:4} is then %
\begin{equation} \label{sen:5}
   \lim_{\epsilon \rightarrow 0}
   \int_0^\infty dr~ r^n \bigl(\frac{1}{r^2} \HU_{\a}\bigr)_\epsilon^2
 = \lim_{a \rightarrow 0} \int_a^\infty dr~ \frac{dr}{r^{4-n}} + \OOO(\epsilon)
 = \lim_{a \rightarrow 0} \frac{a^{n-3}}{3-n} + \OOO(\epsilon),
\end{equation}
because, as $\epsilon \rightarrow 0$, the embedding of $r^{-2}\HU_{\a}(r)$ is $0$ for $r < a$ and $1/r^2$ for $r > a$.  To be rigorous, let us look at $\bigl(r^{-2}\HU_{\a}(r)\bigr)_\epsilon$ more closely.  We have, since $a>0$ and $\epsilon>0$,
\begin{align}
\nonumber 
      \bigl(\frac{1}{r^2} \HU_{a}\bigr)_\epsilon
   & =\lim_{a \rightarrow 0} \int_{\frac{a-r}{\epsilon}}^\infty dz~
                             \frac{\rho(z)}{(r+\epsilon z)^2}\\
\label{sen:6}
   &= \lim_{a \rightarrow 0}
      \frac{1}{r^2} \int_{\frac{a-r}{\epsilon}}^\infty dz~ \rho(z)
      \Bigl(1 - 2\epsilon\frac{z}{r}
            + \dots + (n+1) \bigl( -\epsilon\frac{z}{r}\bigr)^n \Bigr).
\end{align}
Then, if $\epsilon \rightarrow 0$ and $0 < r < a$ the integral tends towards $\int_{\rightarrow +\infty}^{+\infty}$, which vanishes in that limit because  $\rho \in \mathcal{S}$.  On the other hand, if again $\epsilon \rightarrow 0$ but $r>a$ the integral tends towards $\int_{\rightarrow -\infty}^{+\infty}$.  In that case the first term in the development \eqref{sen:6} tends towards $1$, i.e., the normalization \eqref{seq:3}, and the other terms vanish as $\epsilon \rightarrow 0$. Thus, the integrand in \eqref{sen:5} is equal to either $0$ or $r^n/r^4 +  \OOO(\epsilon)$, so that \eqref{sen:5} is proved.

\noindent (ii) Next, the $\HU_{\a}\DUP_{\a}$ term in \eqref{sen:4} is
\begin{equation} \label{sen:7}
  - \lim_{\epsilon \rightarrow 0}
    \int_0^\infty dr~ r^n ~ 2\bigl(\frac{1}{r^2} \HU_{\a}\bigr)_\epsilon
        \bigl(\frac{1}{a} \DUP_{\a}\bigr)_\epsilon
    = - \lim_{a \rightarrow 0} a^{n-3} + \OOO(\epsilon).
\end{equation}
To prove this we start from the identity
\begin{equation} \label{sen:8}
    \Bigr( \bigl(\frac{1}{r^2} \HU_{a}\bigr)_\epsilon^2 \Bigr)'
  = 2 \bigl(\frac{1}{r^2} \HU_{\a}\bigr)_\epsilon ~
    \frac{1}{a} \bigl(\frac{1}{a} \DUP_{\a}\bigr)_\epsilon
  - 2 \bigl(\frac{1}{r^2} \HU_{\a}\bigr)_\epsilon
    2 \bigl(\frac{1}{r^3} \HU_{\a}\bigr)_\epsilon,
\end{equation}
so that the left-hand side of \eqref{sen:7} can be written
\begin{equation} \label{sen:9}
  - \lim_{\epsilon \rightarrow 0} a \int_0^\infty dr~ r^n
    \Bigr( \bigl(\frac{1}{r^2} \HU_{a}\bigr)_\epsilon^2 \Bigr)'
  - \lim_{\epsilon \rightarrow 0} 4a \int_0^\infty dr~ r^n
    \bigl(\frac{1}{r^2} \HU_{\a}\bigr)_\epsilon
    \bigl(\frac{1}{r^3} \HU_{\a}\bigr)_\epsilon,
\end{equation}
which after integrating by part the first integral becomes
\begin{align}
\nonumber
  - \lim_{\epsilon \rightarrow 0} 
    a r^n  \bigl(\frac{1}{r^2} \HU_{a}\bigr)_\epsilon^2 \Bigr|_0^\infty
 &+ \lim_{\epsilon \rightarrow 0} a \int_0^\infty dr~ nr^{n-1}
    \bigl(\frac{1}{r^2} \HU_{a}\bigr)_\epsilon^2\\
\label{sen:10}
 &- \lim_{\epsilon \rightarrow 0} a \int_0^\infty dr~ 4r^n
    \bigl(\frac{1}{r^2} \HU_{\a}\bigr)_\epsilon
    \bigl(\frac{1}{r^3} \HU_{\a}\bigr)_\epsilon.
\end{align}
Now, as $\epsilon \rightarrow 0$, the boundary term is zero because of \eqref{poi:12}, and the $\HU$ terms in the two integrals yield Heaviside functions just like when integrating \eqref{sen:5}.  Therefore, as $\epsilon \rightarrow 0$, we get
\begin{equation} \label{sen:11}
     \lim_{a \rightarrow 0}
      a \int_0^\infty dr~ n r^{n-5}
   - \lim_{a \rightarrow 0}
      a \int_0^\infty dr~ 4 r^{n-5} 
 = - \lim_{a \rightarrow 0} a^{n-3},
\end{equation}
which confirms \eqref{sen:7}.

\noindent (iii) Finally, making the change of variable $r=a + \epsilon z$, the $\DUP_{\a}^2$ term in \eqref{sen:4} is
\begin{align}
\nonumber
    \lim_{\epsilon \rightarrow 0}
    \int_0^\infty dr~ \frac{r^n}{a^2} \frac{1}{\epsilon^2} 
    \rho^2\bigl(\frac{a -r}{\epsilon}\bigr)
 &= \lim_{\epsilon \rightarrow 0}
    \int_{-a/\epsilon}^\infty \frac{dz}{\epsilon} 
    \frac{(a + \epsilon z)^n}{a^2}  
    \rho^2(z)\\
\label{sen:12}
 &= \lim_{\epsilon \rightarrow 0} \Bigl( \frac{a^{n-2}}{\epsilon} M[^2_0]
         + n a^{n-3} M[^2_1] + \OOO(\epsilon) \Bigr),
\end{align}
where $M[^2_0]$ and $M[^2_1]$ are given by \eqref{seq:8}.  Note that $M[^2_1]=0$ when the regularization is even, i.e., if $\rho(-z)=\rho(z)$, in which case the embedding \eqref{seq:4} of the $\delta$-function is even, as is by definition the case of the Dirac measure.

   Adding the three contribution (\ref{sen:5}, \ref{sen:7}, \ref{sen:12}) we get for \eqref{sen:4}
\begin{align}
\nonumber
     \mathcal{M}_n(\a,\epsilon) &=  \int_0^\infty dr~ r^n \vec{E}_\epsilon^2(\vec{r}\,)\\
 \label{sen:13}
       &= \lim_{a \rightarrow 0} \lim_{\epsilon \rightarrow 0}
        \Bigl(  \frac{a^{n-3}}{3-n}  - a^{n-3}
              + \frac{a^{n-2}}{\epsilon} M[^2_0]
              + n a^{n-3} M[^2_1]
              + \OOO(\epsilon)
        \Bigr),
\end{align}  
Then, with $n=2$ and assuming that $\rho$ is even so that $M[^2_1]=0$, the self-energy \eqref{sen:3} is
\begin{equation} \label{sen:14}
 U_{\text{ele}}   = \lim_{a \rightarrow 0} \lim_{\epsilon \rightarrow 0}
   \frac{e^2}{2} \Bigl( \frac{1}{a} - \frac{1}{a}
                      + \frac{1}{\epsilon} M[^2_0] + \OOO(\epsilon) \Bigr)
   = \frac{e^2}{2} \lim_{\epsilon \rightarrow 0}
         \Bigl( \frac{1}{\epsilon} M[^2_0] + \OOO(\epsilon) \Bigr),
\end{equation}
which has the remarkable property to be independent of $a$ because the two $1/a$ contributions cancel {exactly}.  Eq.~\eqref{sen:13} is thus rigorously valid in the limit $a \rightarrow 0$, that is for a \emph{point} electron.

   Summarizing, when the self-energy is calculated in $(\mathcal{C}^\infty)^\mathcal{I}$ rather than in $\mathcal{D}'$, the $\delta(r)/r$ and $\HU(r)/r^2$ terms in the field \eqref{poi:7} interfere in such a way that in the self-energy integral \eqref{sen:3} the mixed term cancels the first term, i.e., the divergent  classical Coulomb-field self-energy \eqref{sen:2}.  The sole contribution to the self-energy comes then from the $\delta^2(r)$ term in \eqref{sen:3}.  This yields the result \eqref{sen:14}, which depends only on $\epsilon$ and on $\rho$ through the value of $M[^2_0]$ given by \eqref{seq:8}, and which may be renormalized to a finite quantity such as the mass of the point charge. 

   We have therefore obtained the physically remarkable result that in the regularization algebra the self-energy of a point-charge is entirely located at the position of the charge, and to order $\OOO(\epsilon)$ solely due to the square of the $\DUP_{\a}(r)$ term in the electric field \eqref{poi:7}, which itself derives form the $\HU_{\a}(r)$ factor in the potential \eqref{poi:4}.

\section{Magnetic dipole: Fields and self-energy}
%================================================
\label{dip:0}

Applying the principles of Sec.~\ref{poi:0} we define the vector potential of a point magnetic dipole as a distribution, and regularize it according to \eqref{seq:2}.  Therefore
\begin{equation}\label{dip:1}
   \vec A_\epsilon(\vec r) \DEF
             \vec{\mu} \times \vec{u} ~
             \bigl(\frac{1}{r^2} \HU_{\a}\bigr)_\epsilon(r),
\end{equation}
where $\vec{u} = \vec{u} (\theta,\varphi)$ is the unit radial vector, and the magnetic moment $\vec{\mu}$ has the dimension of a charge times a length.  The calculation of the magnetic field strength gives
\begin{equation}\label{dip:2}
    \vec H_\epsilon(\vec r) =  \vec\nabla \times \vec A_\epsilon =
        \Bigl( \vec{\mu} + \vec{u}(\vec{\mu}\cdot\vec{u}) \Bigr) h_1(r)
      + \Bigl( \vec{\mu} - \vec{u}(\vec{\mu}\cdot\vec{u}) \Bigr) h_2(r),       \end{equation}
where
\begin{equation}\label{dip:3}
     h_1(r) = \frac{1}{r} \bigl(\frac{1}{r^2} \HU_{\a}\bigr)_\epsilon(r),
   \quad \text{and} \quad
     h_2(r) =  \bigl(\frac{1}{a^2} \DUP_{\a}\bigr)_\epsilon(r)
            - 2\bigl(\frac{1}{r^3}  \HU_{\a}\bigr)_\epsilon(r),     
\end{equation}
which reduces to the classical expression
\begin{equation}\label{dip:4}
    \vec H(\vec r) = 
        \frac{ 3\vec{u}(\vec{\mu}\cdot\vec{u}) - \vec{\mu} }{r^3},
\end{equation}
when $\HU_{\a}$ is replaced by $1$, and $\DUP_{\a}$ by $0$.

Thus, as for the electric field \eqref{poi:7}, there is a $\delta$-function term in the magnetic field \eqref{dip:2}.  In fact, when integrated over 3-space, this $\delta$-function gives the well known finite contribution \cite[p.184]{JACKS1975-}
\begin{equation}\label{dip:5}
      \lim_{{\epsilon \rightarrow 0}} \iiint d^3\Omega ~
       \Bigl( \vec{\mu} - \vec{u}(\vec{\mu}\cdot\vec{u}) \Bigr)
       \bigl(\frac{1}{a^2} \DUP_{\a}\bigr)_\epsilon(r)
     = \frac{8\pi}{3} \vec\mu ,
\end{equation}
which is essential in calculating the hyperfine splitting of atomic $s$-states. Therefore, contrary to the $(\mathcal{C}^\infty)^\mathcal{I}$-expression \eqref{poi:7} of the electric field of a point-charge, in which the $\delta$-term is not directly observable, we have in the magnetic dipolar field \eqref{dip:2} a directly observable $\delta$-like term \cite{GSPON2004D}.

   At this point we could calculate the source current density using Maxwell's equation $ \vec\nabla \times \vec H_\epsilon(\vec r) = \tfrac{4\pi}{c} \vec j_\epsilon(\vec r)$.  But this would lead to a rather complicated formula which we will not explicitly need in this paper.

   Instead, we want to calculate the magnetic self-energy of the point-dipole. Classically, with $\vec{H}$ expressed as \eqref{dip:4}, it is
\begin{equation} \label{dip:6}
 U_{\text{mag}} \DEF \frac{1}{8\pi} \iiint_{\mathbb{R}^3} d^3r~ \vec{H}^2 
 %                   = \frac{1}{2} \int_0^\infty dr~ r^2\frac{e^2}{r^4}
                = \frac{\mu^2}{3}\lim_{r \rightarrow 0} \frac{1}{r^3} = \infty.
\end{equation}
In $(\mathcal{C}^\infty)^\mathcal{I}$, with $\vec{H}$ given by \eqref{dip:2}, we have
\begin{align}
\nonumber
    H_\epsilon^2(\vec r)
 &= \Bigl( \vec{\mu} + \vec{u}(\vec{\mu}\cdot\vec{u})   \Bigr)^2 h_1^2(r)
  +2\Bigl(   {\mu}^2 -        (\vec{\mu}\cdot\vec{u})^2 \Bigr) h_1(r) h_2(r)\\
\label{dip:7}
 &+ \Bigl( \vec{\mu} - \vec{u}(\vec{\mu}\cdot\vec{u})   \Bigr)^2 h_2^2(r).
\end{align}
Then, making the angular integrations, we get
\begin{align}
\label{dip:8}
   \frac{1}{8\pi} \iint d\omega~ H_\epsilon^2(\vec r)
             = \mu^2 \Bigl(  h_1^2(r)
         + \frac{2}{3} h_1(r) \, h_2(r)
         + \frac{1}{3} h_2^2(r) \Bigr).
       \end{align}
Developing this expression we are led to the radial integrals
\begin{align}
\nonumber
 U_{\text{mag}} &=  \mu^2 \int_0^\infty dr~ r^2 
              \frac{1}{r^2} \bigl(\frac{1}{r^2} \HU_{\a}\bigr)_\epsilon^2(r)\\
\nonumber
 &+  \mu^2 \int_0^\infty dr~ r^2 
     \frac{2}{3} \frac{1}{r} \bigl(\frac{1}{r^2} \HU_{\a}\bigr)_\epsilon(r)
                     \Bigl( \bigl(\frac{1}{a^2} \DUP_{\a}\bigr)_\epsilon(r)
                  - 2\bigl(\frac{1}{r^3}  \HU_{\a}\bigr)_\epsilon(r) \Bigr)\\
\label{dip:9}
 &+  \mu^2 \int_0^\infty dr~ r^2 
    \frac{1}{3} \Bigl( \bigl(\frac{1}{a^2} \DUP_{\a}\bigr)_\epsilon(r)
                    - 2\bigl(\frac{1}{r^3}  \HU_{\a}\bigr)_\epsilon(r) \Bigr)^2,
\end{align}
which can be rearranged as
\begin{align}
\label{dip:10}
  U_{\text{mag}} &= \mu^2 \int_0^\infty dr \Bigl(
                    \bigl(\frac{1}{r^2}  \HU_{\a}\bigr)_\epsilon^2
  - \frac{4}{3} r   \bigl(\frac{1}{r^2}  \HU_{\a}\bigr)_\epsilon
                    \bigl(\frac{1}{r^3}  \HU_{\a}\bigr)_\epsilon
  + \frac{4}{3} r^2 \bigl(\frac{1}{r^3}  \HU_{\a}\bigr)_\epsilon^2 \Bigr)\\
\label{dip:11}
 &+ \mu^2 \int_0^\infty dr \Bigl(  
    \frac{2}{3} r   \bigl(\frac{1}{r^2}  \HU_{\a}\bigr)_\epsilon
                    \bigl(\frac{1}{a^2} \DUP_{\a}\bigr)_\epsilon
  - \frac{4}{3} r^2 \bigl(\frac{1}{r^3}  \HU_{\a}\bigr)_\epsilon
                    \bigl(\frac{1}{a^2} \DUP_{\a}\bigr)_\epsilon \Bigr)\\
\label{dip:12}
 &+ \mu^2 \int_0^\infty dr \Bigl(  
    \frac{1}{3} r^2 \bigl(\frac{1}{a^2} \DUP_{\a} \bigr)_\epsilon^2 \Bigr),
\end{align}
where we left the $r$-dependence of $\HU_{\a}$ and $\DUP_{\a}$ implicit.  The technique for calculating these integrals is the same as in Sec.~\ref{sen:0}.  It consists of integrating by parts the mixed terms \eqref{dip:11} so that apart from the $\DUP_{\a}^2$ integral \eqref{dip:12} we are left with integrals of expressions containing products of factors of the form $\bigl(\tfrac{1}{r^n}  \HU_{\a}\bigr)_\epsilon$.  We therefore use the identities
\begin{align}
\label{dip:13}
    r \Bigl( \bigl(\frac{1}{r^2}  \HU_{\a}\bigr)_\epsilon^2\Bigr)'
 &= 2r \bigl(\frac{1}{r^2}  \HU_{\a}\bigr)_\epsilon
       \bigl(\frac{1}{a^2} \DUP_{\a}\bigr)_\epsilon
  - 4r \bigl(\frac{1}{r^2}  \HU_{\a}\bigr)_\epsilon
       \bigl(\frac{1}{r^3}  \HU_{\a}\bigr)_\epsilon,\\
\label{dip:14}    
 a r^2 \Bigl( \bigl(\frac{1}{r^3}  \HU_{\a}\bigr)_\epsilon^2\Bigr)'
 &= 2ar^2 \bigl(\frac{1}{r^3}  \HU_{\a}\bigr)_\epsilon
          \bigl(\frac{1}{a^3} \DUP_{\a}\bigr)_\epsilon
  - 6ar^2 \bigl(\frac{1}{r^3}  \HU_{\a}\bigr)_\epsilon
          \bigl(\frac{1}{r^4}  \HU_{\a}\bigr)_\epsilon,
\end{align}
so that the first integral of \eqref{dip:11} becomes
\begin{align}
\nonumber
    \frac{1}{3}\int_0^\infty dr~
       r \Bigl( \bigl(\frac{1}{r^2}  \HU_{\a}\bigr)_\epsilon^2\Bigr)'
 &+ \frac{4}{3}\int_0^\infty dr~
       r \bigl(\frac{1}{r^2}  \HU_{\a}\bigr)_\epsilon
         \bigl(\frac{1}{r^3}  \HU_{\a}\bigr)_\epsilon\\
\label{dip:15}
 = \frac{1}{3}\bigl(\frac{1}{r^2}  \HU_{\a}\bigr)_\epsilon^2\Bigr|_0^\infty
 - \frac{1}{3}\int_0^\infty dr \bigl(\frac{1}{r^2}  \HU_{\a}\bigr)_\epsilon^2
&+ \frac{4}{3}\int_0^\infty dr~
       r \bigl(\frac{1}{r^2}  \HU_{\a}\bigr)_\epsilon
         \bigl(\frac{1}{r^3}  \HU_{\a}\bigr)_\epsilon,
\end{align}
and the second one of \eqref{dip:11}
\begin{align}
\nonumber
   -\frac{2a}{3}\int_0^\infty dr~
       r^2 \Bigl( \bigl(\frac{1}{r^3}  \HU_{\a}\bigr)_\epsilon^2\Bigr)'
 &- 4a\int_0^\infty dr~
      r^2 \bigl(\frac{1}{r^3}  \HU_{\a}\bigr)_\epsilon
          \bigl(\frac{1}{r^4}  \HU_{\a}\bigr)_\epsilon\\
\label{dip:16}
 = -\frac{2a}{3}r^2\bigl(\frac{1}{r^3}  \HU_{\a}\bigr)_\epsilon^2\Bigr|_0^\infty
 + \frac{4a}{3} \int_0^\infty dr~
      r   \bigl(\frac{1}{r^3}  \HU_{\a}\bigr)_\epsilon^2
&- 4a\int_0^\infty dr~
      r^2 \bigl(\frac{1}{r^3}  \HU_{\a}\bigr)_\epsilon
          \bigl(\frac{1}{r^4}  \HU_{\a}\bigr)_\epsilon.
\end{align}
In the limit  $\epsilon \rightarrow 0$ the boundary terms in (\ref{dip:15}--\ref{dip:16}) are zero, and, adding all the $\HU_{\a}$ contributions in \eqref{dip:10} and (\ref{dip:15}--\ref{dip:16}), we obtain to order $\OOO(\epsilon)$
\begin{align}
\label{dip:17}
 \Bigl(1 - \frac{4}{3} + \frac{4}{3} - \frac{1}{3} + \frac{4}{3} \Bigr)
  \int_a^\infty \frac{dr}{r^4}
  + \Bigl(\frac{4}{3} - 4 \Bigr)
  a\int_a^\infty \frac{dr}{r^5} = \frac{2}{3a^3} - \frac{2}{3a^3} =0.
 \end{align}
Therefore, as in the case of the electric self-energy, the contributions from the $\HU_{\a}^2$ and $\HU_{\a}\DUP_{\a}$ terms, i.e., from \eqref{dip:10} and \eqref{dip:11}, cancel each other to order $\OOO(\epsilon)$, and to that order the sole non-zero contribution to the magnetic self-energy comes from the $\DUP_{\a}^2$ term, i.e., using \eqref{sen:12},
\begin{align}
\nonumber
  U_{\text{mag}} &= \mu^2 \int_0^\infty dr \Bigl(  
    \frac{1}{3} r^2 \bigl(\frac{1}{a^2} \DUP_{\a} \bigr)_\epsilon^2 \Bigr)\\
\label{dip:18}
   &= \frac{1}{3} \lim_{a \rightarrow 0} \lim_{\epsilon \rightarrow 0}
     \frac{\mu^2}{a^2} \Bigl( \frac{1}{\epsilon} M[^2_0]
                        + \frac{2}{a} M[^2_1] + \OOO(\epsilon) \Bigr),
\end{align}
where $M[^2_1]=0$ if the regularization is even.

\section{Electron singularity: Self-force and stability}
%=======================================================
\label{sfo:0}

Starting in this section we calculate the basic dynamical properties of a classical electron at rest defined as a pole-dipole singularity of the Maxwell field, that is as an electric-pole of charge $e$ and a magnetic-dipole of moment $\vec{\mu}$ located at one same point, taken as the origin of the coordinate system.  

   To do this we assume that these properties derive solely from the electromagnetic fields generated by that singularity, so that, for instance, the total self-energy of the electron is
\begin{align}
\label{sfo:1}
  U_{\text{electron}} = \frac{1}{8\pi} \iiint_{\mathbb{R}^3} d^3r~
      (\vec{E}^2+\vec{H}^2) = U_{\text{ele}} + U_{\text{mag}},
\end{align}
where $U_{\text{ele}}$ and $U_{\text{mag}}$ are given by \eqref{sen:14} and \eqref{dip:18}.

Similarly, the total self-force, is given by
\begin{align}
\label{sfo:2}
  \vec{F}_{\text{electron}} =  \iiint_{\mathbb{R}^3} d^3r~
      (\varrho\vec{E} + \vec{j} \times \vec{H}),
\end{align}
where $\vec{E}$ and $\varrho$ are given by \eqref{poi:7} and  by \eqref{poi:10}, and  $\vec{H}$ by \eqref{dip:2} while $\vec{j}$ may be derived from \eqref{dip:2} by means of Maxwell's equation $ \vec\nabla \times \vec H_\epsilon(\vec r) = \tfrac{4\pi}{c} \vec j_\epsilon(\vec r)$.

   However, as is discussed in \cite[Sec.\,8]{GSPON2008B}, it turns out that the total self-force \eqref{sfo:2} is zero. Indeed, since the electron is strictly point-like, the forces which potentially tend to its disassembly compensate each other exactly at one \emph{single} point after angular integration:  The electron is therefore stable and Poincar\'e compensating stresses are not required.  

   But this necessitates that all the terms in \eqref{sfo:2} are finite when  $\epsilon \neq 0$ and $a \neq 0$, enabling the radial and angular integrations to be interchanged without their order affecting the result.  As in \cite[Sec.\,8]{GSPON2008B} we will verify that this is indeed the case for the electric part of the radial self-force --- the calculation of the magnetic part being similar but more lengthy.  Hence, we calculate the Lorentz self-force on a electric point-charge.   Because the potential is just a function of $r$, the electric field can be written $\vec{E}(\vec{r}\,) = E(r) \vec{u}$, where $\vec{u}$ is the unit vector in the direction of $\vec{r}$, and the charge density $\varrho(\vec{r}\,)=\varrho(r)$.  Thus, by definition, the electric self-force is
\begin{equation} \label{sfo:3}
 \vec{F}_{\text{ele}} \DEF \iiint_{\mathbb{R}^3} d^3r~ \varrho(\vec{r}\,)\vec{E}(\vec{r}\,) 
                 =  \iint d\omega \vec{u}(\theta,\varphi) ~F_{\text{r}},
\end{equation}
where the radial force is given by
\begin{equation} \label{sfo:4}
        F_{\text{r}} = \int_0^\infty dr~ r^2 \varrho(r) E(r).
\end{equation}
Of course, as $\vec{u}$ is just the unit radius vector, integrating over $4\pi$ solid angle yields a zero total self-force if the radial and angular integrals can be interchanged, which is the case if $F_{\text{r}}$ is finite. 

In the classical theory this radial force is the integral
\begin{equation} \label{sfo:5}
 \   F_{\text{r}} =  \int_0^\infty dr~ r^2
                           \frac{e}{r^2} \delta(r) \frac{e}{r^2} 
                  =  e^2\lim_{r \rightarrow 0} \frac{1}{r} = \infty,
\end{equation}
so that the electron is absolutely unstable since the self-force is outwards-directed and infinite.  In distribution theory, with the Coulomb field and charge-density expressed as \eqref{poi:8} and \eqref{poi:11}, the result is the same as \eqref{sfo:5} with $r$ replaced by the parameter $a$.

In regularized distribution theory we could calculate \eqref{sfo:4} using for $\vec{E}$ and $\varrho$ expressions \eqref{poi:7} and \eqref{poi:10}, but this would lead to laborious calculations since ten different terms would have to be integrated.  It is easier to exploit the fact that $\vec{E}_\epsilon$ and $\varrho_\epsilon$, i.e., the regularizations of the distributions $\vec{E}$ and $\varrho$, are actually $\mathcal{C}^\infty$ functions so that one can use integration by parts to put \eqref{sfo:4} in a more convenient form.  In view of this we note that with $E = |\vec{E}_\epsilon(\vec{r}\,)|$ we can write $\varrho=E' + 2 E/r$, so that the radial integration in \eqref{sfo:4} becomes
\begin{equation} \label{sfo:6}
 F_{\text{r}} =  \int_0^\infty dr~ r^2 E \Bigl(E' + \frac{2}{r} E \Bigr).
\end{equation}
Then, we integrate by parts using the identity
\begin{equation} \label{sfo:7}
      E \Bigl(E' + \frac{2}{r} E \Bigr)
   = \frac{1}{2r^4} \Bigl(2r^4 E E' + 4r^3 E^2 \Bigr)
   = \frac{1}{2r^4} \Bigl( \bigl(r^2 E^2\bigr)^2 \Bigr)',
\end{equation}
which is well defined at $r=0$ because of \eqref{poi:12}. Thus
\begin{equation} \label{sfo:8}
 F_{\text{r}} =  \frac{1}{2} r^2E^2 \Bigr|_0^\infty
              + \int_0^\infty dr~ r E ^2,
\end{equation}
where the boundary term is zero on account of \eqref{poi:12}.  Therefore,
\begin{equation} \label{sfo:9}
    F_{\text{r}}  =  \int_0^\infty dr~ r \vec{E}_\epsilon^2(\vec{r}\,)
      = \mathcal{M}_1(\a,\epsilon),
\end{equation}  
and the radial self-force is given by the first moment of $\vec{E}_\epsilon^2$, i.e., from \eqref{sen:13}, assuming that the regularizing function $\rho$ is even,
\begin{equation} \label{sfo:10}
 F_{\text{r}}   = \lim_{a \rightarrow 0} \lim_{\epsilon \rightarrow 0}
                  \Bigl( \frac{1}{2a^2} - \frac{1}{a^2}
                       + \frac{1}{a\epsilon} M[^2_0] + \OOO(\epsilon) \Bigr).
\end{equation}
Which shows that, in the self-force, the mixed term is over-compensating the $\text{H}^2$ term, i.e., 
\begin{equation} \label{sfo:11}
 F_{\text{r}}   = \lim_{a \rightarrow 0} \lim_{\epsilon \rightarrow 0}
                  \Bigl( \frac{1}{a\epsilon} M[^2_0]
                       - \frac{1}{2a^2} + \OOO(\epsilon) \Bigr)
                  \geq 0,
\end{equation}
because $\epsilon \ll a$.

In conclusion, $F_{\text{r}}$ is finite in $(\mathcal{C}^\infty)^\mathcal{I}$ if both $a$ and $\epsilon$ are non-zero, which is necessary for the calculations to make sense in  $(\mathcal{C}^\infty)^\mathcal{I}$.  Thus, under these conditions, the total electric self-force is zero after angular integration.

\section{Electron singularity: Hidden momentum and spin}
%=========================================================
\label{ele:0}

In this section we calculate the self-momentum and self-angular-momentum of an electron singularity defined as an electric monopole of charge $e$ located at the same position as a magnetic dipole of momentum $\vec{\mu}$, i.e., the origin of the coordinate system.  These momenta are defined like the momentum and angular-momentum of an electromagnetic field, i.e.,
\begin{align}
\label{ele:1}
 \vec{P} &\DEF \frac{1}{4\pi c} \iiint_{\mathbb{R}^3} d^3r~
              \vec{E} \times \vec{H},\\
\label{ele:2}
 \vec{S} &\DEF \frac{1}{4\pi c} \iiint_{\mathbb{R}^3} d^3r~
              \vec{r} \times (\vec{E} \times \vec{H}\,),
\end{align}
but $\vec{P}$ is called `hidden momentum' and $\vec{S}$ `spin' when the corresponding sources are at rest, as the case for an electron singularity attached to the origin of the coordinate system.

  For example, a system consisting of a point magnetic-dipole of moment $\vec{m}$ located at $\vec{r} = 0$, and a point charge $q$ positioned at $\vec{r} = \vec{a}$, has a non-zero hidden momentum \cite{GSPON2007C}
\begin{eqnarray}\label{ele:3}
    \vec{P}
      =  \dfrac{q}{a^3} \vec{m} \times \vec{a},
\end{eqnarray}
which tends towards infinity as $1/a^2$ when the point charge approaches the position of the point dipole.  We therefore expect that the hidden momentum of an electron singularity could also be infinite, unless it is zero as in the case of the self-force.  To find out we have have to make a detailed analysis.

Taking for $\vec{E}$ and $\vec{H}$ expressions \eqref{poi:7} and \eqref{dip:2}, we easily find, since $\vec{u}\times\vec{u}=0$,
\begin{align}
\nonumber
     \vec{E} \times \vec{H} =  e \vec{\mu} \times \vec{u} ~
    &\Bigl( \bigl(\frac{1}{r^2} \HU_{\a}\bigr)_\epsilon(r)
           - \frac{1}{a}\bigl( \DUP_{\a}\bigr)_\epsilon(r) \Bigr)\\
\label{ele:4}
    \times
    &\Bigl(2\bigl(\frac{1}{r^3}  \HU_{\a}\bigr)_\epsilon(r)
        - \frac{1}{r}   \bigl(\frac{1}{r^2} \HU_{\a}\bigr)_\epsilon(r)
        - \frac{1}{a^2} \bigl(             \DUP_{\a}\bigr)_\epsilon(r) \Bigr).
\end{align}
Separating angular and radial integrations, \eqref{ele:1} and \eqref{ele:2} are then
\begin{align}
\label{ele:5}
 \vec{P} &= \frac{1}{4\pi c} \iint d\omega~ 
              e \vec{\mu} \times \vec{u} ~R_2(\a,\epsilon),\\
\label{ele:6}
 \vec{S} &= \frac{1}{4\pi c} \iint d\omega~
              \vec{u} \times (e \vec{\mu} \times \vec{u}\,) ~R_3(\a,\epsilon),
\end{align}
where, developing and rearranging \eqref{ele:4}, the radial integrals are
\begin{align}
\label{ele:7}
        R_n(\a,\epsilon) &=  \int_0^\infty dr~  \Bigl(
      2r^n \bigl(\frac{1}{r^2}  \HU_{\a}\bigr)_\epsilon
            \bigl(\frac{1}{r^3}  \HU_{\a}\bigr)_\epsilon
   - r^{n-1}\bigl(\frac{1}{r^2}  \HU_{\a}\bigr)_\epsilon^2 \Bigr)\\
\label{ele:8}
     &+  \int_0^\infty dr~  \Bigl(
     - \frac{2r^n}{a} \bigl(\frac{1}{r^3}  \HU_{\a}\bigr)_\epsilon
                                   \bigl( \DUP_{\a}\bigr)_\epsilon
             +\frac{ar^{n-1}-r^n}{a^2}
            \bigl(\frac{1}{r^2}  \HU_{\a}\bigr)_\epsilon
                         \bigl( \DUP_{\a}\bigr)_\epsilon \Bigr)\\
\label{ele:9}
     &+  \int_0^\infty dr~  \Bigl(
         \frac{r^n}{a^3}\bigl( \DUP_{\a}\bigr)_\epsilon^2 \Bigr),
\end{align}
in which the $r$-dependencies of $\HU_{\a}$ and $\DUP_{\a}$ are implicit. Of course, since \eqref{ele:5} is odd in the substitution $\vec{u} \rightarrow \vec{u}$, its angular integral is zero, whereas \eqref{ele:6} being even its angular part is non-zero.  Indeed, separating the angular and radial integrals and writing $P_r=R_2(\a,\epsilon)$ and $S_r=R_3(\a,\epsilon)$, we have 
\begin{align}
\label{ele:10}
 \vec{P} &= P_r \, \vec{P}_\omega, \qquad \text{where} \qquad
                   \vec{P}_\omega = 0,\\
\label{ele:11}
 \vec{S} &= S_r \, \vec{S}_\omega, \qquad \text{where} \qquad
                   \vec{S}_\omega = \frac{2e}{3c} \vec{\mu}.
\end{align}
But we want to study the radial integrals independently of the fact that the angular integrals are zero or not. 

To integrate \eqref{ele:9} we use the identities
\begin{align}
\label{ele:12}
  a^2r^n \Bigl( \bigl(\frac{1}{r^3}  \HU_{\a}\bigr)_\epsilon^2\Bigr)'
 &= \frac{2r^n}{a} \bigl(\frac{1}{r^3}  \HU_{\a}\bigr)_\epsilon
                                \bigl( \DUP_{\a}\bigr)_\epsilon
  - 6a^2r^n \bigl(\frac{1}{r^3}  \HU_{\a}\bigr)_\epsilon
            \bigl(\frac{1}{r^4}  \HU_{\a}\bigr)_\epsilon,\\
\nonumber    
 \frac{ar^{n-1}-r^n}{2} \Bigl( \bigl(\frac{1}{r^2}  \HU_{\a}\bigr)_\epsilon^2\Bigr)'
 &= \frac{ar^{n-1}-r^n}{a^2} \bigl(\frac{1}{r^2}  \HU_{\a}\bigr)_\epsilon
                                          \bigl( \DUP_{\a}\bigr)_\epsilon\\
\label{ele:13}
 &- 2(ar^{n-1}-r^n) \bigl(\frac{1}{r^2}  \HU_{\a}\bigr)_\epsilon
                    \bigl(\frac{1}{r^3}  \HU_{\a}\bigr)_\epsilon,
\end{align}
so that the first integral in \eqref{ele:8} becomes
\begin{align}
\nonumber
   - \int_0^\infty dr~
     a^2r^n \Bigl( \bigl(\frac{1}{r^3}  \HU_{\a}\bigr)_\epsilon^2\Bigr)'
  &- \int_0^\infty dr~
    6a^2r^n \bigl(\frac{1}{r^3}  \HU_{\a}\bigr)_\epsilon
            \bigl(\frac{1}{r^4}  \HU_{\a}\bigr)_\epsilon\\
\nonumber
 = -a^2r^n \bigl(\frac{1}{r^3}  \HU_{\a}\bigr)_\epsilon^2\Bigr|_0^\infty
  &+ \int_0^\infty dr~
      a^2nr^{n-1} \bigl(\frac{1}{r^3}  \HU_{\a}\bigr)_\epsilon^2\\
\label{ele:14}
  &- \int_0^\infty dr~
6a^2r^n \bigl(\frac{1}{r^3}  \HU_{\a}\bigr)_\epsilon
            \bigl(\frac{1}{r^4}  \HU_{\a}\bigr)_\epsilon,
\end{align}
and the second one
\begin{align}
\nonumber
     \int_0^\infty dr
       \frac{ar^{n-1}-r^n}{2}
       \Bigl( \bigl(\frac{1}{r^2}  \HU_{\a}\bigr)_\epsilon^2\Bigr)'
  &+ \int_0^\infty dr~
   2(ar^{n-1}-r^n) \bigl(\frac{1}{r^2}  \HU_{\a}\bigr)_\epsilon
                   \bigl(\frac{1}{r^3}  \HU_{\a}\bigr)_\epsilon\\
\nonumber
 =     \frac{ar^{n-1}-r^n}{2}
       \bigl(\frac{1}{r^2}  \HU_{\a}\bigr)_\epsilon^2\Bigr|_0^\infty
  &- \int_0^\infty dr~
     \frac{a(n-1)r^{n-2}-nr^{n-1}}{a^2}
     \bigl(\frac{1}{r^2}  \HU_{\a}\bigr)_\epsilon^2\\
\label{ele:15}
  &+ \int_0^\infty dr~
   2(ar^{n-1}-r^n) \bigl(\frac{1}{r^2}  \HU_{\a}\bigr)_\epsilon
                   \bigl(\frac{1}{r^3}  \HU_{\a}\bigr)_\epsilon.
\end{align}
In the limit  $\epsilon \rightarrow 0$ the boundary terms in (\ref{ele:14}--\ref{ele:15}) are zero, and, adding all the $\HU_{\a}$ contributions in \eqref{ele:7} and (\ref{ele:14}--\ref{ele:15}), we obtain in the same limit
\begin{align}
\label{ele:16}
  \int_a^\infty dr\Bigl(
  \frac{n-2}{2}   r^{n-5}
- \frac{n-5}{2}a  r^{n-6}
+   (n-6)      a^2r^{n-7} \Bigr)
 = \frac{3-n}{n-4}a^{n-4}.
 \end{align}
It remains to calculate the $\DUP_{\a}^2$-term in \eqref{ele:9} in the same limit, i.e., using \eqref{sen:12} and assuming that the regularizing function $\rho$ is even
\begin{align}
\label{ele:17}
  \lim_{\epsilon \rightarrow 0} \int_0^\infty dr~  \Bigl(
         \frac{r^n}{a^3}\bigl( \DUP_{\a}\bigr)_\epsilon^2 \Bigr)
   =  \lim_{a \rightarrow 0} \lim_{\epsilon \rightarrow 0}
      \Bigl( \frac{a^{n-3}}{\epsilon} M[^2_0] + \OOO(\epsilon) \Bigr),
\end{align}
so that adding \eqref{ele:16} we have
\begin{align}
\label{ele:18}
  \lim_{\epsilon \rightarrow 0} R_n(\a,\epsilon) 
   =  \lim_{a \rightarrow 0} \lim_{\epsilon \rightarrow 0}
      \Bigl( \frac{a^{n-3}}{\epsilon} M[^2_0]
            -\frac{n-3}{n-4}a^{n-4} + \OOO(\epsilon) \Bigr).
\end{align}
Therefore, the radial integrals in (\ref{ele:10}--\ref{ele:11}) are
\begin{align}
\label{ele:19}
     P_r &=  \lim_{a \rightarrow 0}
             \lim_{\epsilon \rightarrow 0} \Bigl(
             \frac{1}{a\epsilon} M[^2_0] -\frac{1}{2a^2}
                          + \OOO(\epsilon) \Bigr),\\
\label{ele:20}
     S_r &=  \lim_{\epsilon \rightarrow 0}
             \Bigl( \frac{1}{\epsilon} M[^2_0] + \OOO(\epsilon) \Bigr).
\end{align}
The radial component of the self-momentum has thus the same form as the radial component of the self-force \eqref{sfo:11}.  On the other hand the spin is independent on $a$ and a pure $\delta^2$ integral just like the electric self-energy \eqref{sen:14}, with the difference that here the $\delta^2$ comes from the product of the $\delta$-contributions in $\vec{E}$ and $\vec{H}$.

In conclusion, combining (\ref{ele:10}--\ref{ele:11}) and  (\ref{ele:19}--\ref{ele:20}) we get
\begin{align}
\label{ele:21}
     \vec{P} &= 0,\\
\label{ele:22}
     \vec{S} &=  \frac{2e}{3c} \vec{\mu} \lim_{\epsilon \rightarrow 0}
                 \Bigl( \frac{1}{\epsilon} M[^2_0]+ \OOO(\epsilon) \Bigr).
\end{align}
Therefore, while the hidden-momentum is zero (a necessity for the stability of the electron) the spin is, to order $\OOO(\epsilon)$, a $\delta^2$-quantity located at the position of the electron just like its electric self-energy \eqref{sen:14}, as well as its magnetic self-energy \eqref{dip:18} when the regularization function $\rho$ is even.

\section{Discussion}
%===================
\label{dis:0}

In this section we compare our results to those obtained in Ref.\,\cite{GSPON2008B}.  To do this we compare the three main numbers obtained, the self-energies $U_{\text{ele}}$ and $U_{\text{mag}}$ whose sum corresponds to the electromagnetic self-mass $m$ of the electron, i.e.,
\begin{align}
\label{dis:1}
   mc^2 = U_{\text{ele}} + U_{\text{mag}},
\end{align}
and the electromagnetic self-angular-momentum $S= |\vec{S}|$, i.e., the spin of the electron.\footnote{Let us stress again that what is done here is a purely classical calculation, and that we do not take into account the constraints that may arise from general principles such as Wigner-Poincar\'e invariance, etc. Thus, neither the mass nor the spin are `quantized,' and all results should be interpreted in the same spirit as, for example, the angular-momentum radiated by the multipolar field of an accelerated classical electron \cite[p.\,750--751]{JACKS1975-}.}

\def\CI{(\mathcal{C}^\infty)^\mathcal{I}}
\def\CO{\mathcal{G}}

   In the present paper, working in the sequence algebra $(\mathcal{C}^\infty)^\mathcal{I}$ and assuming an even regularizing function $\rho$, we have found, i.e., Eqs.\,\eqref{sen:14}, \eqref{dip:18}, and \eqref{ele:22}, that
\begin{align}
\label{dis:2}
  U_{\text{ele}}(\CI) &= \lime \frac{1}{2} \frac{M_{\rho}[^2_0]}{\epsilon}
                              ~ e^2 + \OOO(\epsilon),\\
\label{dis:3}
  U_{\text{mag}}(\CI) &= \limea \frac{1}{3} \frac{M_{\rho}[^2_0]}{\epsilon}
                              \frac{\mu^2}{a^2} + \OOO(\epsilon),\\
\label{dis:4}
               S(\CI) &= \lime \frac{2}{3} \frac{M_{\rho}[^2_0]}{\epsilon}
                              \frac{e \mu}{c} + \OOO(\epsilon),
\end{align}
which mean that these equalities are first order approximations in $\epsilon$ as $\epsilon \rightarrow 0$.

    On the other hand, in Ref.\,\cite{GSPON2008B}, working in the Colombeau algebra $\mathcal{G}$ and assuming an even mollifier $\eta$,  we have found that
\begin{align}
\label{dis:5}
  U_{\text{ele}}(\CO) &= \lime  \frac{1}{2} \frac{M_{\eta}[^2_0]}{\epsilon}
                              ~ e^2 + \OOO(\epsilon),\\
\label{dis:6}
  U_{\text{mag}}(\CO) &= \limea \frac{1}{3} \frac{M_{\eta}[^2_0]}{\epsilon}
                              \frac{\mu^2}{a^2} + \OOO(\epsilon),\\
\label{dis:7}
               S(\CO) &= \lime \frac{2}{3} \frac{M_{\eta}[^2_0]}{\epsilon}
                              \frac{e \mu}{c} + \OOO(\epsilon),
\end{align}
which correspond to Eqs.\,(10.1), (10.2), and (10.3) of Ref.\,\cite{GSPON2008B}, and in which we have made explicit that these expressions are also  $\OOO(\epsilon)$ approximations when taking Eq.\,(5.33) of Ref.\,\cite{GSPON2008B} into account.

  Therefore, comparing (\ref{dis:2}--\ref{dis:4}) to (\ref{dis:5}--\ref{dis:7}), it looks as if there were no difference between working in $\CI$ or in $\CO$ --- apart for  $M[^2_0]$ having a different value in the two sets when $\rho\neq \eta$.  But this is not the case:  Each of these formulas is the sum of several terms, and making the contribution arising from the $\delta^2$-term in each of them explicit reveals a huge \emph{qualitative} difference.

  Indeed, let us define $\delta_{\epsilon,\chi}(r) \DEF \epsilon^{-1}\chi(-r/\epsilon)$. Then, as is easily verified, Eq.~\eqref{sen:12} of the present paper, in which $\delta^2_{\epsilon,\rho}$ is integrated in $\CI$, is (replacing $\rho$ by $\eta$) equivalent to Eq.\,(5.33) of Ref.\,\cite{GSPON2008B}, in which $(\UPS')^2 = \delta^2_{\epsilon,\eta} $ is integrated in $\CO$.  That is, in both frameworks,
\begin{align} 
 \label{dis:8}
  \int_0^\infty \delta^2_{\epsilon,\chi}(r) ~ F(r) ~dr
 = \limea  M_{\chi}[^2_0]\frac{F(a)}{\epsilon} + \OOO(\epsilon),
\end{align}
provided the regularizer/mollifier is even, as we always assumed.  Thus, looking at the details of the calculations made in the present paper and in Ref.\,\cite{GSPON2008B}, one finds
\begin{align}
\label{dis:9}
  U_{\text{ele}}(\CI) &= \lime \frac{1}{2} e^2
                         \int_0^\infty \delta^2_{\epsilon,\rho}(r) ~dr
                          + \OOO(\epsilon),\\
\label{dis:10}
  U_{\text{mag}}(\CI) &= \limea \frac{1}{3} \frac{\mu^2}{a^2} 
                         \int_0^\infty \delta^2_{\epsilon,\rho}(r) ~dr
                               + \OOO(\epsilon),\\
\label{dis:11}
               S(\CI) &= \lime  \frac{e \mu}{c}
                         \int_0^\infty \delta^2_{\epsilon,\rho}(r) ~dr
                              + \OOO(\epsilon),
\end{align}
whereas
\begin{align}
\label{dis:12}
  U_{\text{ele}}(\CO) &= \lime \frac{1}{2} e^2
                         \int_0^\infty \delta^2_{\epsilon,\eta}(r) ~dr
                          + \OOO(\epsilon^q),\\
\label{dis:13}
  U_{\text{mag}}(\CO) &= \limea \frac{1}{3} \frac{\mu^2}{a^2} 
                         \int_0^\infty \delta^2_{\epsilon,\eta}(r) ~dr
                               + \OOO(\epsilon^q),\\
\label{dis:14}
               S(\CO) &= \lime  \frac{e \mu}{c}
                         \int_0^\infty \delta^2_{\epsilon,\eta}(r) ~dr
                              + \OOO(\epsilon^q),
\end{align}
where the $\OOO(\epsilon^q)$ contributions with $\epsilon \in \mathcal{I}$ and $q \in \mathbb{N}$ as large as we please can be ignored for all practical purposes, so that Eqs.\, (\ref{dis:12}--\ref{dis:14}) are actually \emph{equalities}  in $\mathcal{G}$ --- as was stressed in the conclusion of Ref.\,\cite{GSPON2008B}.

   In conclusion, in a Colombeau algebra, the electron's electromagnetic self-energies and the self-angular-momentum, i.e., mass and spin, are exact integrals of $\delta^2$-functions.  Moreover, as was recalled in the introduction, the product of smooth functions in a Colombeau algebra is not `regularization dependent,' as is the case in the sequence algebra $\CI$.  This means that embedding classical electrodynamics in a Colombeau algebra, and allowing for truly point-like singularities as well as nonlinear operations,  are mathematically consistent and physically meaningful.

%\section{Conclusion}
%===================
\label{con:0}

%\section{Acknowledgments}
%========================

\section{References}
%====================
\label{biblio:0}

\end{document}